\pdfoutput=1
\newcommand*{\ATLASLATEXPATH}{}
\documentclass[PAPER, atlasdraft=true, texlive=2016, UKenglish, british, PAPER, cernpreprint]{\ATLASLATEXPATH atlasdoc}
\DeclareOldFontCommand{\rm}{\normalfont\rmfamily}{\mathrm}
 
\usepackage[title]{appendix}
\usepackage[minimal, backend=biber]{\ATLASLATEXPATH atlaspackage}
 
\newcommand{\papertype}{letter}

\usepackage[articletitle=true,showdoi=false]{\ATLASLATEXPATH atlasbiblatex}
\usepackage{\ATLASLATEXPATH atlascontribute}
\usepackage{\ATLASLATEXPATH atlasphysics}
\usepackage{\ATLASLATEXPATH atlasheavyion}
\usepackage{\ATLASLATEXPATH atlasunit}
\usepackage{\ATLASLATEXPATH atlasprocess}
\usepackage{ZPbPb_paper-defs}
\usepackage{authblk}
 
\graphicspath{{logos/}{figures/}}
 
\usepackage{ANA-HION-2018-10-PAPER-defs}
\usepackage{array}
\newcolumntype{P}[1]{>{\centering\arraybackslash}p{#1}}

\AtlasTitle{$Z$ boson production in Pb+Pb collisions at $\sqrt{s_{\textrm{NN}}}$~=~5.02~TeV~measured by the ATLAS experiment}
 
\AtlasAbstract{
The production yield of $Z$ bosons is measured in the electron and muon decay channels in Pb+Pb collisions at $\sqrt{s_{\textrm{NN}}}$~=~5.02~TeV~with the ATLAS detector.  Data from the 2015 LHC run corresponding to an integrated luminosity of 0.49~$\textrm{nb}^{-1}$~are used for the analysis. The $Z$ boson yield, normalised by the total total number of minimum-bias events and the mean nuclear thickness function, is measured as a function of dilepton rapidity and event centrality.  The measurements in Pb+Pb collisions are compared with similar measurements made in proton--proton collisions at the same centre-of-mass energy. The nuclear modification factor is found to be consistent with unity for all centrality intervals. The results are compared with theoretical predictions obtained at next-to-leading order using nucleon and nuclear parton distribution functions. The normalised $Z$ boson yields in Pb+Pb collisions lie 1--3$\sigma$ above the predictions. The nuclear modification factor measured as a function of rapidity agrees with unity and is consistent with a next-to-leading-order QCD calculation including the isospin effect.
}
 
\author{The ATLAS Collaboration}
\PreprintIdNumber{CERN-EP-2019-182}
 
\AtlasRefCode{HION-2018-10}

\AtlasJournal{Phys.\ Lett.\ B.}

\AtlasCoverCommentsDeadline{XX 2019}
 
\AtlasCoverAnalysisTeam{Mirta Duman\v{c}i\'{c}, Alexander Milov, Fabrice Balli, Zvi Citron}
 
\AtlasCoverEdBoardMember{Peter~Kluit~(chair), Gideon~Bella, Nikolopoulos~Konstantinos}

\AtlasCoverEgroupEditors{atlas-HION-2018-10-editors@cern.ch}
 
\AtlasCoverEgroupEdBoard{atlas-HION-2018-10-editorial-board@cern.ch}

\hypersetup{pdftitle={ATLAS document},pdfauthor={The ATLAS Collaboration}}
\usepackage{siunitx}
\usepackage{booktabs}
\usepackage{dcolumn}
\newcolumntype{i}[1]{S[table-format=#1.0]}
\newcolumntype{d}[1]{S[table-format=#1.1]}
\usepackage{pbox}
 
\begin{document}
 
\maketitle

\section{Introduction
}
The measurement of electroweak (EW) boson production is a key part of the heavy-ion (HI) physics programme at the Large Hadron Collider (LHC). Isolated photons and heavy vector bosons, \Zboson and \Wboson, are powerful tools to probe the initial stages of HI collisions. 
After being created at the initial stage of the collision in high-momentum exchange processes, \Zboson and \Wboson bosons decay much faster than the timescale of the medium's evolution. Their leptonic decay products are generally understood to not be affected by the strong interaction; hence they carry information about the initial stage of the collision and the partonic structure of the nuclei. \par
 
Measurements performed by the ATLAS and CMS experiments with \Zboson and \Wboson bosons decaying leptonically show that production rates of these non-strongly interacting particles are proportional to the amount of nuclear overlap, quantified by the mean nuclear thickness function,~\avgTaa~\cite{Aad:2012ew,Aad:2014bha,Chatrchyan:2011ua,Chatrchyan:2014csa,Chatrchyan:2012nt, Aad:2019sfe}. Results obtained with isolated high-energy photons~\cite{Aad:2015lcb,Chatrchyan:2012vq} are also consistent with this observation.
 
The transverse momentum and rapidity distribution of \Zboson bosons and the pseudorapidity distribution of muons originating from \Wboson bosons measured in \PbPb collisions at $\sqn=2.76$~\TeV\ have been found to be generally consistent with perturbative quantum chromodynamics (pQCD) calculations of nucleon--nucleon collisions scaled by \avgTaa. Production of $Z$ bosons in \PbPb collisions was found to be consistent with next-to-leading-order (NLO) pQCD calculations that do not include nuclear modifications in the treatment of parton distribution functions (PDFs). However, some nuclear modification of PDFs could not be excluded within the precision of the existing \PbPb measurements~\cite{Aad:2012ew,Aad:2014bha,Aad:2015lcb}. The recent ALICE result at $\sqn=5.02$~\TeV\ shows better agreement with nPDF calculations at forward rapidities~\cite{Acharya:2017wpf}.
 
On the other hand, the study of asymmetric \pPb collisions at $\sqn=5.02$~\TeV\ shows that including nuclear modifications of PDFs gives a substantially better description of the data than using a free proton PDF.  This is seen by comparing the \Zboson boson cross section in \pPb collisions with pQCD calculations~\cite{Aaij:2014pvu, Aad:2015gta, Khachatryan:2015pzs, Alice:2016wka} and recently also the $W$ boson cross section at $\sqn=5.02$~\TeV\ and $\sqn=8.16$~\TeV~\cite{Alice:2016wka, Sirunyan:2019dox}. In addition, studies of \Zboson bosons differentially in \pPb centrality demonstrate that they are a sensitive test of the Glauber model description of nuclear geometry~\cite{Aad:2015gta}.
 
This \papertype~presents results on \Zboson boson production yield measurement in the \Zmm and \Zee decay channels in \PbPb collisions at $\sqn = 5.02$~\TeV\ with the ATLAS detector at the LHC. The data sample was collected in November 2015 and corresponds to an integrated luminosity of 0.49~\inb. The observables under study are the yield of produced \Zboson bosons in the fiducial kinematic region defined by detector acceptance and lepton kinematics normalised to the number of minimum-bias events, measured differentially in rapidity and event centrality.  The \PbPb data are compared with pQCD calculations, and the nuclear modification factor is measured relative to previously measured \pp\ cross sections~\cite{pp_paper_zw2019}.

\newcommand{\AtlasCoordFootnote}{
ATLAS uses a right-handed coordinate system with its origin at the nominal interaction point (IP)
in the centre of the detector and the $z$-axis along the beam pipe.
The $x$-axis points from the IP to the centre of the LHC ring,
and the $y$-axis points upwards.
Cylindrical coordinates $(r,\phi)$ are used in the transverse plane,
$\phi$ being the azimuthal angle around the $z$-axis.
The pseudorapidity is defined in terms of the polar angle $\theta$ as $\eta = -\ln \tan(\theta/2)$.}
 
\section{ATLAS detector\label{sec:atlas2}}
 
The ATLAS detector~\cite{PERF-2007-01} covers nearly the entire solid angle\footnote{\AtlasCoordFootnote}
around the collision point.
It consists of an inner tracking detector surrounded by a thin superconducting solenoid,
electromagnetic and hadronic calorimeters, and a muon spectrometer incorporating three
large superconducting toroid magnets.
 
The inner-detector system is immersed in a \SI{2}{\tesla} axial magnetic field and
provides charged-particle tracking in the range $|\eta| < 2.5$.
The high-granularity silicon pixel detector covers the vertex region and typically provides four measurements per track, the first hit being in the insertable B-layer~\cite{ATLAS-TDR-2010-19, Abbott:2018ikt} in operation since 2015. It is followed by the silicon microstrip tracker, which usually provides eight measurements per track. These silicon detectors are complemented by the transition-radiation tracker, which enables radially extended track reconstruction up to $|\eta| = 2.0$.
 
The calorimeter system covers the pseudorapidity range $|\eta|<4.9$. Within the region $|\eta|< 3.2$, electromagnetic (EM) calorimetry is provided by barrel and
endcap high-granularity liquid-argon (LAr) sampling calorimeters,
with an additional thin LAr presampler covering $|\eta| < 1.8$,
to correct for energy loss in material upstream of the calorimeters.
Hadronic calorimetry is provided by the scintillator-tile calorimeter,
segmented into three barrel structures within $|\eta| < 1.7$, and two LAr hadronic endcap calorimeters.
The forward calorimeter (FCal) is a LAr sampling calorimeter located on either side of the
interaction point. It covers $3.1 < |\eta| < 4.9$ and each half is composed of one EM and two hadronic
sections. The FCal is used to
characterise the centrality of Pb+Pb collisions as described below. Finally, zero-degree calorimeters
(ZDC) are situated at large pseudorapidity, $|\eta| > 8.3$, and are primarily sensitive to spectator neutrons. \par
 
The muon spectrometer comprises separate trigger and
high-precision tracking chambers measuring the deflection of muons in a magnetic field generated by superconducting air-core toroids.
The precision chamber system covers the region $|\eta| < 2.7$ with three layers of monitored drift tubes,
complemented by cathode-strip chambers in the forward region, where the background is highest.
The muon trigger system covers the range $|\eta| < 2.4$ with resistive-plate chambers in the barrel, and thin-gap chambers in the endcap regions.
 
A two-level trigger system is used to select
events of interest for recording~\cite{Aaboud:2016leb}.  The level-1 (L1) trigger is implemented in
hardware and uses a subset of the detector information to reduce the event
rate. The subsequent, software-based
high-level trigger (HLT) selects events for recording. Both the electron and muon event selection used in this analysis combine L1 and HLT decision algorithms.
 
\section{Data sets and event selection \label{sec:Datasets}}
 
All of the analysed data were recorded in periods with stable beam, detector, and trigger operations. Candidate events are required to have at least one primary vertex reconstructed from the inner-detector tracks. In addition, a trigger selection is applied, requiring a muon or an electron candidate above a \pT threshold of 8~\GeV\ or 15~\GeV, respectively.  The electron-trigger candidate is further required to satisfy a set of loose criteria for the electromagnetic shower shapes~\cite{Aad:2019wsl}. The trigger algorithm implements an event-by-event estimation and subtraction of the underlying-event contribution to the transverse energy deposited in each calorimeter cell~\cite{Aad:2012vca}. For both the electron and muon candidates, further requirements are applied to suppress electromagnetic background contributions, as described in Section~\ref{sec:background}. 
\par
 
Muon candidates reconstructed offline must satisfy $\pT>20$~\GeV\ and $|\eta|$~<~2.5 and pass the requirements of `medium' identification optimised
for 2015 analysis conditions~\cite{Aad:2016jkr}. Offline selected electron candidates are required to have $\pT>20$~\GeV~and $|\eta|<2.47$, although candidates within the transition region between barrel and endcap calorimeters ($1.37 < |\eta|< 1.52$) are rejected. In addition, `loose' likelihood-based identification is applied, developed for the Pb+Pb data conditions and based on a general strategy described in Ref.~\cite{Aaboud:2019ynx}. \par 
 
Events with a \Zboson boson candidate are selected by requiring exactly two opposite-charge muons or electrons, at least one of which is matched to a lepton selected at trigger level. The dilepton invariant mass must satisfy the requirement $66 < \mll < 116$~\GeV. A total of 5347 $Z$ boson candidates are found in the muon channel and 4047 in the electron channel. 
\par

In order to estimate the geometric characteristics of HI collisions, it is common to classify the events according to the amount of nuclear overlap in the collision. The quantity used to estimate the collision geometry is called the `collision centrality'.
The centrality determination is based on the total transverse energy measured by both FCal detectors in each event, \sumfcalet.
This quantity is then mapped to geometric quantities, such as the average number of participating nucleons, \avgNpart, and the mean nuclear thickness function, \avgTAA, which quantifies the amount of nuclear overlap in a centrality class and is evaluated using a Glauber calculation~\cite{Miller:2007ri, Aaboud:2018quy}.
The mapping is based on specific studies of an event sample without additional Pb+Pb collisions within the same or neighbouring bunch crossings (pile-up) collected with minimum-bias (MB) triggers.
A special treatment is employed for events in the 20\% most peripheral interval, where diffractive and photonuclear processes contribute significantly to the MB event sample. This requires extrapolating from the total number of MB events in this region and employing a special requirement on the \Zboson boson event topology, as described in Section~\ref{sec:background}. Table \ref{table:centrality} summarises the relationship between centrality, \avgNpart, and \avgTAA as calculated with Glauber MC v2.4~\cite{photon_jet}, which incorporates nuclear densities averaged over protons and neutrons. The total number of MB events in the 0--80\% centrality interval is $(2.99\pm0.04)\times10^{9}$, which is then distributed in different centrality intervals according to their size. The quoted uncertainty on the number of MB events includes variations on the \sumfcalet value corresponding to the 0--80\% centrality interval estimated with the Glauber model. This sample is obtained by selecting events passing MB triggers and excluding the events with a pile-up contribution, where the total sampled integrated luminosity corresponds to the signal selection~\cite{Aaboud:2018quy}. \par
 
\begin{table}[h!]
\centering
\caption{Centrality intervals and their corresponding geometric quantities with systematic uncertainties, from Ref.~\cite{photon_jet}.}
\begin{tabular}{ l|cc c l|cc}
Centrality [\%]  & \avgNpart& \avgTaa [mb$^{-1}$]  & & Centrality [\%]  & \avgNpart& \avgTaa [mb$^{-1}$]  \\
\cline{1-3} \cline{5-7}
0--2\%     &  $399.0\pm1.6$ & $28.30\pm0.25$   & & 20--25\% &  $205.6\pm2.9$ & $9.77\pm0.18$ \\
2--4\%     &  $380.2\pm2.0$ & $25.47\pm0.21$  & & 25--30\% &  $172.8\pm2.8$ &$7.50\pm0.17$\\
4--6\%     &  $358.9\pm2.4$ & $23.07\pm0.21$  & & 30--40\% &  $131.4\pm2.6$ &$4.95\pm0.15$   \\
6--8\%     &  $338.1\pm2.7$ & $20.93\pm0.20$  & & 40--50\% &  $~~87.0\pm2.4$ &$2.63\pm0.11$   \\
8--10\%   &  $317.8\pm2.9$ & $18.99\pm0.19$   & & 50--60\% &  $~~53.9\pm2.0$ &$1.28\pm0.07$   \\
10--15\% &  $285.2\pm2.9$ & $16.08\pm0.18$  & & 60--80\% &  $~~23.0\pm1.3$ &$0.39\pm0.03$   \\
15--20\% &  $242.9\pm2.9$ & $12.59\pm0.17$  & & 80--100\% &  $~~~~4.80\pm0.36$ &$0.052\pm0.006$   \\
\hline
&				&		             & & 0--100\% &  $114.0\pm1.1$ &$5.61\pm0.06$
 
\end{tabular}
\label{table:centrality}
\end{table}

Simulated samples of Monte Carlo~(MC) events are used to evaluate the selection efficiency for signal events and the contribution of several background processes to the analysed data set. All of the samples were produced with the
\textsc{Geant4}-based simulation~\cite{Agostinelli:2002hh, SOFT-2010-01} of the ATLAS detector. Dedicated efficiency and calibration studies with data are used to derive correction factors to account for residual differences between experiment and simulation.
 
The processes of interest containing \Zboson bosons were generated with the \Powheg-\textsc{Box} v1 MC program~\cite{Nason:2004rx,Frixione:2007vw,Alioli:2010xd,Alioli:2008gx} interfaced to the \Pythia\ 8.186 parton shower model~\cite{Sjostrand:2007gs}. The CT10 PDF set~\cite{Gao:2013xoa} was used in the matrix element, while the CTEQ6L1 PDF set~\cite{Pumplin:2002vw} was used with the AZNLO~\cite{STDM-2012-23} set of generator-parameter values (tune) for the modelling of non-perturbative effects in the initial-state parton shower. The \Photos++ v3.52 program~\cite{Davidson:2010ew} was used for final-state photon radiation in EW processes. \par
 
A sample of top-quark pair~(\ttbar) production was generated with the \Powheg-\textsc{Box} v2 generator, which uses NLO matrix-element calculations~\cite{Frixione:2007nw} together with the CT10f4 PDF set~\cite{Lai:2010vv}. 
The parton shower, fragmentation and underlying event in nucleon--nucleon collisions were simulated using \Pythia\ 6.428~\cite{Sjostrand:2006za} with the CTEQ6L1 PDF set and the corresponding Perugia 2012 tune (P2012)~\cite{Skands:2010ak}. The top-quark mass was set to 172.5~\GeV. The EvtGen v1.2.0 program~\cite{Lange:2001uf} was used
to model 
bottom and charm hadron decays for all versions of \Pythia.  
The total \Zboson boson and top-quark yields in MC samples are normalised using the results of NLO QCD calculations.\par
 
The signal MC samples were produced with different nucleon--nucleon combinations ($pp$, $pn$, $nn$) weighted to reflect the isospin composition of lead nuclei. For lead, $A=208$ and $Z=82$, so all samples with two neutrons have a weight of $([A-Z]/Z)^{2}=$36.7\%, events with two protons have a weight of 15.5\% and the unlike nucleon combinations ($pn$, $np$) each have weights of 23.9\%.

Once produced, the simulated events were overlaid with MB events taken during the \PbPb run. The overlay of data events was done such that the MC simulation accurately reflects detector occupancy conditions present in the \PbPb run.  The MB events used for the overlay were sampled such that the centrality distribution, based on the total transverse energy deposited in the forward calorimeters, approximates that of \Zboson boson events, which are generally biased to more-central collisions. The simulated events were finally reconstructed by the standard ATLAS reconstruction software.  
\par
\section{Analysis}
\label{sec:analysis_big}
\subsection{Measurement procedure 
}
The differential \Zboson boson production yield per MB event is measured within a fiducial phase space defined by $\ptl>20$~\GeV, $|\etal|<2.5$ and $66<\mll<116$~\GeV.
The yields in both the electron and muon channel are calculated using
\begin{equation}
N_{Z}^\mathrm{fid} = \frac{N_{Z} - B_{Z}}{ \epsilon^{Z}_{\textrm{trig}} \cdot \CZ } \,,
\label{eq:Zxsecfidemu}
\end{equation}
where $N_{Z}$ and $B_{Z}$ are the number of selected events in data and the expected number of background events, respectively,
and $\epsilon^{Z}_{\textrm{trig}}$ is the trigger efficiency per \Zboson boson candidate measured in data and described in Section~\ref{sec:corrections}. 
 
A correction for the reconstruction efficiency, momentum resolution and the final-state radiation effects is applied with the bin-by-bin correction factor \CZ\, which is obtained from MC  simulation as
\begin{equation}
\nonumber
\CZ = \frac{N^\mathrm{MC, sel}_{Z}}{N^\mathrm{MC, fid}_{Z}}\,.
\end{equation}
Here, $N^\mathrm{MC, sel}_{Z}$ is the number of events passing the signal selection at the detector level. The number of selected events is corrected for measured differences between data and simulation in lepton reconstruction and identification efficiencies. The denominator $N^\mathrm{MC, fid}_{Z}$ is computed by applying the fiducial phase-space requirements to the generator-level leptons originating from \Zboson boson decays. The measurement is corrected for QED final-state radiation effects by using the generator-level lepton kinematics before photon radiation.
 
The value of \CZ\ in the fiducial acceptance averaged over all centralities is determined from MC simulation after reweighting as explained in Section~\ref{sec:Datasets}. It is $0.659$ and $0.507$ in the \zmm\ and \zee\ decay channels, respectively. The uncertainty due to the size of the simulated sample is at the level of 0.1\% for each decay channel and is not the dominant uncertainty.

The rapidity, momentum and centrality dependence of \CZ\ is calculated from the simulation as
\begin{equation}
\CZ (\pt,y,\Sigma E_{\textrm{T}}^{\textrm{FCal}}) = F(\pt,y)G(y, \sumfcalet),
\label{eq:Z_eff}
\end{equation}
where $F(\pt,y)$ is the centrality-averaged efficiency calculated per $y$ and \pt interval of the dilepton system and $G(y, \sumfcalet)$ is a parabolic parameterisation of a correction factor accounting for the centrality and rapidity dependences of the efficiency. In each rapidity bin, the factor $G$ is obtained from a fit of the ratio of the efficiency in a particular centrality bin to the value averaged over all possible centrality values.

Nuclear modification is quantified by measuring the ratio of the \Zboson boson production rate, scaled by the mean nuclear thickness function, to the \Zboson boson production cross section in \pp collisions, and is called the nuclear modification factor:
\begin{equation}
\RAA(y) = \frac{1}{\avgTaa\Nevt}\frac{\mathrm{d}N_{\footnotesize\PbPb}^{\footnotesize Z} / \mathrm{d}y}{\mathrm{d}\mathrm{\sigma}_{\footnotesize\pp}^{\footnotesize Z} / \mathrm{d}y}, \notag
\end{equation}
where \avgTaa is the nuclear thickness function in a given centrality class, $(1/N_{\mathrm{evt}}) \mathrm{d} N_{\footnotesize\PbPb}^{\footnotesize Z}/\mathrm{d}y$ is the differential yield of \Zboson bosons per inelastic MB event measured in \PbPb collisions and $\mathrm{d}\mathrm{\sigma}_{\footnotesize\pp}^{\footnotesize Z} / \mathrm{d}y$ is the differential \Zboson boson cross section measured in \pp collisions~\cite{pp_paper_zw2019}. A deviation from unity in \RAA indicates the nuclear modification of the observable. The value of \RAA is expected to be greater than unity by about 2.5\%, based on MC simulation, due to the higher \Zboson boson production cross section in proton--neutron and neutron--neutron interactions which are present in \PbPb collisions and amount to 84.5\% of the total hadronic cross section. This is later referred to as the `isospin effect' and is not accounted for in the definition of \RAA.

\subsection{Background determination\label{sec:background}}
 
There are two background source categories studied in this analysis. The first includes the same background sources that are studied in \pp collisions~\cite{pp_paper_zw2019} and the second includes additional background sources specific to the \PbPb collision system. 
 
Background contributions in the first category are expected from \ztt, top-quark pair production and multi-jet events. The first two contributions are evaluated from dedicated simulation samples, whereas the multi-jet background contribution is derived using a data-driven approach. The \ztt\ background is found to be 0.05\% of all signal candidates in the muon channel and 0.06\% in the electron channel. The top-quark background amounts to 0.08\% in the muon channel and 0.05\% in the electron channel. The background contribution from $W$ boson decays and $W$+jet production is found to be negligible. \par
 
\begin{figure}[tbp]
\centering
\includegraphics[width=0.49\textwidth]{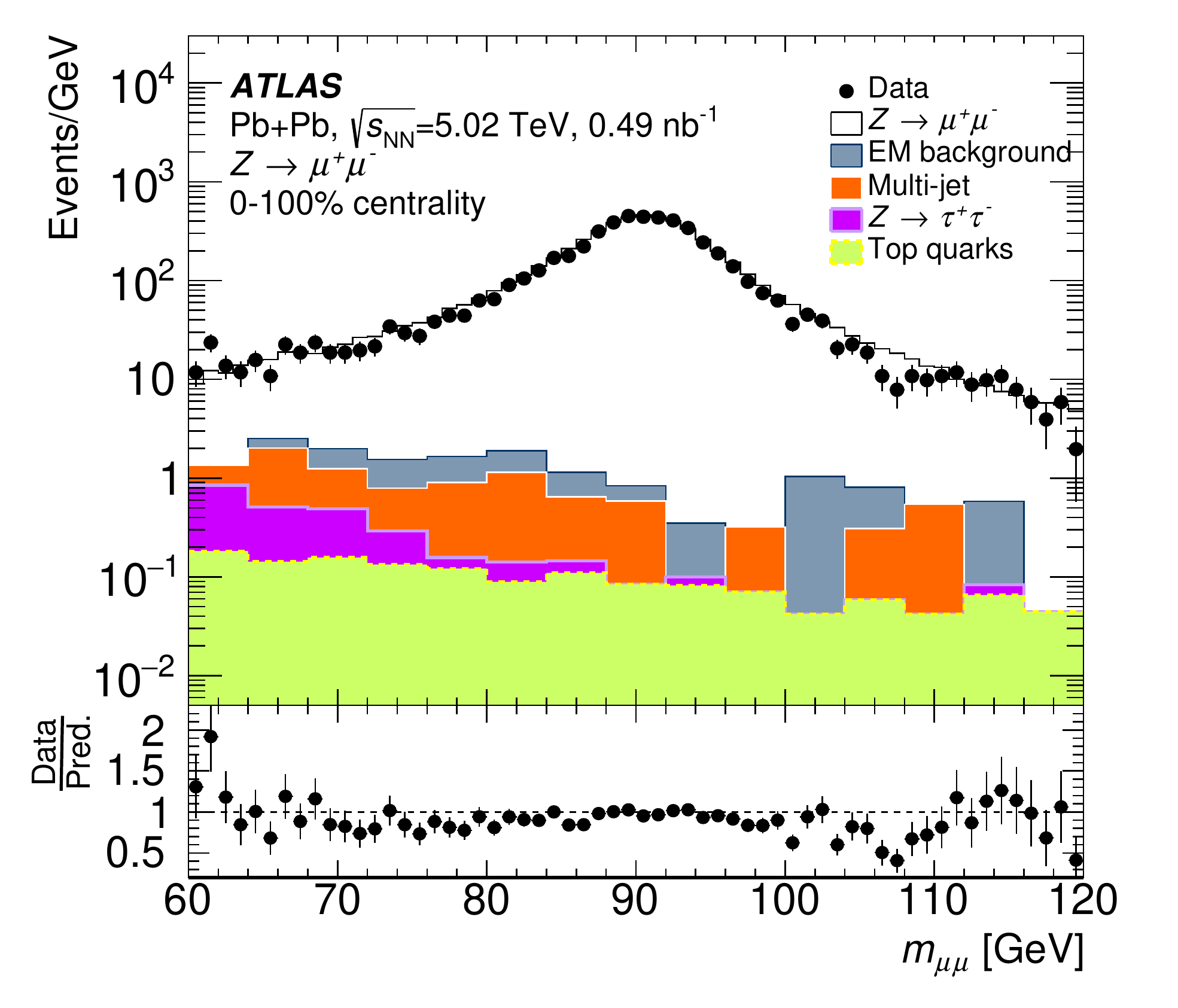}
\includegraphics[width=0.49\textwidth]{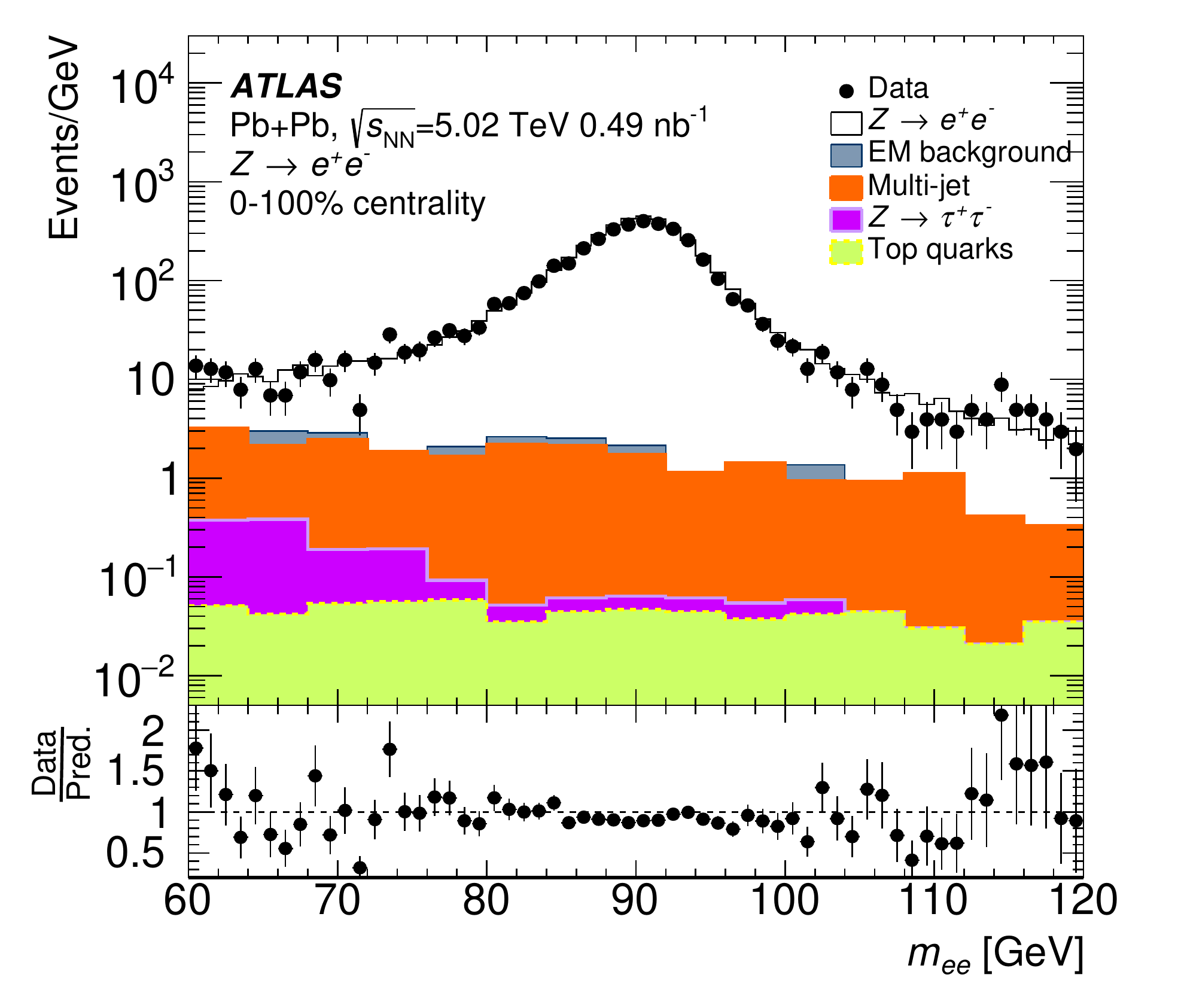}
\caption{Centrality-integrated detector-level invariant mass distribution of (left) dimuon and (right) dielectron pairs together with the \ztt, top quark, multi-jet and EM background contributions. Only the statistical uncertainties of the data are shown. \label{fig:mass_channels}}
 
\end{figure}

The multi-jet background originates from jets, misidentified hadrons and, in the electron channel, from converted photons. In the muon channel, its contribution is estimated from the distribution of the same-sign \Zboson boson candidates in rapidity and centrality. Due to the low charge misidentification rate in the muon spectrometer, their invariant mass distribution does not exhibit a peak in the \Zboson boson mass region. In the muon channel this background amounts to 0.5\% of all signal candidates. In the electron channel, there is a significant contribution from charge misidentification, fakes and conversions. The electron same-sign pairs therefore cannot be used to estimate the multi-jet background. This contribution to the selected event sample in the electron channel is estimated using a background template obtained from the data in \Zboson boson rapidity and event centrality.  The template is derived from a subset of the signal sample that corresponds to electrons from jets, i.e.\ electrons with a very poor reconstruction quality. It is normalised to the number of same-sign data candidates in the low-mass region of $60<m_{ee}<70$~GeV after the signal MC subtraction. Due to the small number of signal candidates satisfying this condition, same-sign electron pairs with the same kinematic requirements are also added to this background template. The shape of the obtained multijet template is shown in Figure~\ref{fig:mass_channels}. This background amounts to 2\% of all signal candidates in the electron channel. \par

The background contributions specific to Pb+Pb come from two main sources. The first is due to pile-up when more than a single \PbPb collision is recorded simultaneously or in a nearby bunch crossing. The second is the production of additional \Zboson boson candidates by photon-induced reactions produced by the intense electromagnetic fields generated by the colliding ions (below referred to as `electromagnetic background').
Pile-up distorts the transverse energy measured in the FCal and causes reconstructed \Zboson bosons to be assigned to an incorrect centrality interval.
Pile-up events from other collisions in the same bunch crossing (in-time pile-up) increase the \sumfcalet, shifting the \Zboson boson candidate to a more-central interval. Alternatively, if a pile-up collision precedes the trigger event (out-of-time pile-up), its contribution to the \sumfcalet can be negative, due to the time response of the electronic signal shapers used in the calorimeters~\cite{Aaboud:2018ugz}. In this case, the $Z$ boson candidate is shifted to a more-peripheral interval.
Both processes depend on the instantaneous luminosity during data taking. At any time during the HI run the number of hadronic interactions per bunch crossing was less than 0.01. To preserve the accuracy of the total yield measurement, no pile-up removal procedure was applied to the selected events. However, due to the fact that the \Zboson boson production scales linearly with \avgTAA~\cite{Aad:2012ew}, the increase in the FCal transverse energy in in-time pile-up events transfers candidates from less populated to more-populated centrality intervals, thus having a very small effect and changing the average number of counts in the most central collisions by an insignificant amount. Contrary to that, the reduction in the \sumfcalet transfers out-of-time pile-up events from more-populated to less-populated centrality intervals, thus making a larger relative contribution to the more-peripheral events.
The effect has been studied using several independent data-driven and simulation-based approaches. The largest contribution to the most peripheral 80--100\% centrality interval due to this type of the pile-up is less than 2\%, i.e.\ less than one count, and is significantly less in any other centrality range.
 
A non-negligible relative contribution to the dileptons in the $Z$ boson mass range in peripheral centrality intervals is expected from electromagnetic background sources.
On the other hand, the expected rate of signal events in those peripheral centrality bins is low. Two photon-induced processes are expected to contribute to the background: photon--photon scattering, $\gamma \gamma \rightarrow \ell^{+} \ell^{-}$~\cite{Bertulani:1987tz, Krauss:1997vr, Baur:2008hn} and photon--nucleus scattering $\gamma +A \rightarrow Z\rightarrow \ell^{+}\ell^{-}$~\cite{article_photoproduction}. Although measurements of exclusive high-mass dilepton production have been performed in $pp$ and Pb+Pb collisions by ATLAS~\cite{Aad:2015bwa} and in $pp$ collisions by CMS~\cite{Chatrchyan:2011ci} a photo-nuclear $Z$ boson production has not yet been observed in HI collisions.
When the impact parameter of the photon-induced processes is larger than twice the nuclear radius such processes are referred to as ultra-peripheral collisions, and in these they are not obscured by hadronic interactions.
Both physics processes are characterised by large rapidity gaps on one or both sides of the detector (regions with no particle production recorded in the detector), which are used in this analysis to measure and subtract these backgrounds. The rapidity gap estimation is implemented using a similar technique as developed in Ref.~\cite{ Aad:2012pw}.

In the 50--100\% (peripheral) centrality intervals, there is an additional requirement of a ZDC signal coincidence in order to suppress the electromagnetic background contributions. The energy measured in either detector is required to be at least 1~TeV, corresponding to 40\% of the energy deposition of a single neutron.
Without using any ZDC coincidence requirement in the event selection, 34 events with a rapidity gap greater than 2.5 units are found in the two decay channels. Since the estimated number of hadronic \Zboson boson candidates with such a gap is below 0.05, all of these events are considered to be produced by photon-induced reactions and are removed from the sample. Events without gaps can have a photon-induced dilepton pair as well, if the rapidity gap is filled by particles originating from a simultaneous nucleon--nucleon interaction. These events would appear in the centrality intervals defined by the \sumfcalet deposition from the hadronic interaction.

Following Ref.~\cite{Aaboud:2018eph}, photon-induced reactions occurring simultaneously with hadronic collisions can be identified using both the angular and momentum correlations of final-state dilepton pairs.
One variable used to quantify these correlations is the dilepton acoplanarity, defined as $\alpha \equiv 1-|\phi^{+}-\phi^{-}|/{\pi}$, where $\phi^{\pm}$ are the azimuthal angles of the two opposite-charge leptons. The same observable is used in this analysis to extract the contribution of photon-induced reactions to the measured $Z$ boson production.
Based on the MC signal simulation and measurements in the 0--50\% centrality interval, ($13.3\pm0.4$)\% of \Zboson boson candidates produced by hadronic collisions have acoplanarity below 0.01. On the other hand, among the 34 events with a rapidity gap greater than 2.5 units which were rejected from the sample as pure photon-induced events, 26 are found to have $\alpha < 0.01$, corresponding to a fraction of 76\%. This demonstrates that the acoplanarity is sensitive to photon-induced reactions in the \Zboson boson invariant-mass region. This allows the photon-induced background to be estimated in all centrality intervals by comparing the number of \Zboson boson candidates in a given centrality interval with the number of candidates with $\alpha < 0.01$.

It is estimated that in the 80--100\% centrality interval, besides the events with a large rapidity gap, $7\pm3$ out of 28 remaining candidates originate from photon-induced reactions. In the 60--80\% interval this number is $15\pm5$ out of 182, and in the 50--60\% interval it is $18\pm8$ out of 258, where the uncertainty includes statistical and systematic uncertainties in the background, but not in the number of event candidates. 
In more-central collisions, the method of estimating the photon-induced background is limited by the statistical precision of the sample and this contribution is consistent with zero. 
Rapidity distributions in centrality intervals are normalised to the estimated number of signal events and an additional systematic error is assigned in each interval to account for the EM background subtraction procedure. \par

Figure~\ref{fig:mass_channels} shows the invariant mass distribution of backgrounds from all considered sources for both decay channels integrated in event centrality.
The EM background shape shown in Figure~\ref{fig:mass_channels} is derived from the events containing large rapidity gaps and is normalised to the total estimated number of background candidates quoted above.  \par

\subsection{Detector performance corrections\label{sec:corrections}}
 
After subtracting background contributions, the number of \Zboson boson candidates is corrected for the trigger efficiency and detection efficiency, according to Eq.~(\ref{eq:Zxsecfidemu}). All the correction factors are derived directly from the current data set used in the analysis, with the exception of the lepton momentum calibration corrections that are derived from \pp collision data, and extrapolated to the Pb+Pb dataset conditions. The trigger efficiency per reconstructed $Z$ candidate $\epsilon^{Z}_{\textrm{trig}}$ is derived from the efficiency of the single-lepton trigger $\epsilon_{\ell}$ via the relation $\epsilon^{Z}_{\textrm{trig}} = 1-(1-\epsilon_{\ell_{1}})(1-\epsilon_{\ell_{2}})$, where the indices refer to the two leptons forming the candidate pair. In order to obtain $\epsilon^{Z}_{\textrm{trig}}$ as a function of the dilepton \pt and $y$ which is further applied as a correction per dilepton candidate, kinematic distributions of the decay products are taken from MC simulation.
 
Muon and electron trigger efficiencies are measured using a tag-and-probe method~\cite{Aaboud:2016leb, Aad:2016jkr, Aaboud:2019ynx} as a function of $\eta$ and $\phi$.
The tag lepton is required to be reconstructed with high quality and very low probability of background contamination and to be matched to a lepton selected at trigger level. The probe lepton, satisfying the analysis reconstruction and identification requirements, is paired to it to give an invariant mass in the range $66 < \mll < 116$~\GeV. 
The background contribution to this measurement is estimated from the number of same-sign pairs and amounts to 0.8\% and 3.5\% in the muon and electron channels, respectively.

The single-muon trigger efficiency in the endcap region of the detector~($1.05<|\eta|<2.4$) is measured to be around 85\%, and in the barrel region~($|\eta|<1.05$) it varies between 60\% and 80\%. A significant dependence of the efficiency on the muon azimuthal angle $\phi$ was measured and thus the trigger efficiency correction is derived as a function of both $\phi$ and $\eta$. The single-electron trigger efficiency is measured to be around 95\% in the endcap region of the calorimeter~($1.52<|\eta|<2.47$) and it increases slightly to 97\% in the barrel~($|\eta|<1.37$). A significant dependence on the electron \pt was measured, and the efficiency rises from 85\% to 97\% in the range from 20 to 100~\GeV\ integrated over $\eta$. The single-electron trigger efficiency is thus derived as a function of \pt and $\eta$. 
The average $\epsilon^{Z}_{\textrm{trig}}$ is ($94.2\pm0.2$)\% in the muon channel and ($99.74\pm0.03$)\% in the electron channel, constant with the event centrality within 1\%.

Selection requirements including the muon reconstruction and identification are imposed on muon candidates used in the analysis. The efficiency of the selection criteria is
measured using a tag-and-probe method in \zmm\ events~\cite{Aad:2016jkr} and compared with simulation.
Ratios of the efficiencies determined in data and simulation are applied as scale factors (SF) to correct the simulated events.
Since the measured efficiencies are found to have negligible dependence on the muon momentum in the selected kinematic region and a very weak centrality dependence, the SF are evaluated only as a function of muon $\eta$. The centrality dependence of the SF is taken into account in the evaluation of systematic uncertainties.
The combined reconstruction and identification efficiency for medium-quality muons typically exceeds 84\% in both the data and simulation with good agreement between the two estimates. The largest difference is observed in the endcap region ($|\eta|>1.8$). 
Ultimately, the SF values for three different $\eta$ regions are as follows:  $0.97\pm0.01$ for $|\eta|<0.8$, $0.99\pm0.01$ for $0.8<|\eta|<1.8$ and $1.04\pm0.02$ for $|\eta|>1.8$. \par

Electron candidates used for the analysis are required to satisfy selection criteria related to reconstruction and identification. The efficiency of the selection is measured using a tag-and-probe method in \zee\ events, as described in Ref.~\cite{Aaboud:2019ynx}, and compared with simulation to derive electron scale factors.  Measurements are performed as a function of the electron $\eta$, \pt and event centrality. 
The electron reconstruction and identification efficiency is measured to be typically 70\% in the endcap ($|\eta|>1.52$) with good agreement between the data and the simulation. The SF is measured to be 1\% away from unity with a precision of 3\% in that region. In the barrel region ($|\eta|<1.37$) the efficiency is measured to be around 80\% while in the MC simulation the efficiency reaches 85\%. Therefore, in this region a significant SF is applied, measured with a precision of 3--5\%. \par

The lepton momentum scale and resolution corrections are derived using the \pp signal MC samples and are applied to the simulation for both the electrons and muons.
For the reconstructed muons, these corrections are derived as a function of the muon $\eta$ and $\phi$~\cite{Aad:2016jkr}.
The correction factors are chosen such that they minimise the $\chi^2$ between the muon-pair invariant mass distributions
in data and simulation. The energy scale of reconstructed electrons is corrected by applying to the data a per-electron correction factor. The momentum scale correction factors are derived from a comparison of the electron-pair invariant mass between simulation and data. \par

\section{Systematic uncertainties}
\label{sec:systematics}

In both channels, the trigger efficiency is derived from the tag-and-probe results in the data. The statistical limitation of the measured sample determines the uncertainty associated with the trigger correction.  Although the uncertainties in each bin are relatively large in the muon trigger efficiency, after propagating this uncertainty to the dimuon efficiency, where only one of the muons is required to fire the trigger, the total uncertainty is quite small, between 0.2\% and 0.5\% when derived as a function of centrality and 0.1--0.2\% when derived as a function of $Z$ boson rapidity.
The uncertainty is propagated using MC pseudo-experiments and the uncertainties in the linear fit coefficients of the trigger efficiency as a function of centrality. In the electron channel this uncertainty is at most 0.5\%. \par
 
The NLO cross section of the background samples of \ztt~and top-quark production is varied by 10\% to derive the corresponding uncertainties~\cite{pp_paper_zw2019}. However, in both decay channels the multi-jet background dominates the uncertainty contribution. In the muon channel, the multijet background contribution is varied by 10\%, which produces 0.01--0.1\% uncertainty in rapidity bins and up to 0.2\% uncertainty in centrality bins. In the electron channel, the multi-jet template normalisation is varied by 20\%, which corresponds to the level of statistical uncertainty in the number of same-sign candidates in the low-mass control region. The overall contribution to the systematic uncertainty is about 0.5\% in rapidity bins and 0.5--2\% in centrality bins. \par
 
In the 50--100\% centrality interval, the uncertainty in subtracting the photon-induced background is evaluated by considering two sources. The first source is the compatibility between the acoplanarity cut efficiencies for hadronic \Zboson boson production evaluated from data and simulation. An uncertainty of 0.4\% accounts for this difference. For the second source, the uncertainty in the background rejection efficiency of the acoplanarity cut is evaluated from the candidates with large rapidity gaps. This uncertainty has two contributions. One is the statistical uncertainty of the event sample with large rapidity gaps, which amounts to 7\%. The other contribution is due to the observed differences between the acoplanarity distributions for electrons and muons, which amounts to about 8\%.
In the 0--50\% centrality interval, where the background subtraction is not performed, an uncertainty of 0.4\%, evaluated from the difference between the data and simulation acoplanarity distributions, is introduced to account for a possible residual EM background contribution.  \par
 
Uncertainties in the determination of lepton reconstruction and identification efficiency scale factors, as well as the parameterisation of the centrality dependence of the total correction affect the measurements through the correction factors \CZ. \par
 
In the muon channel, the scale factors in the three $\eta$ regions described in Section~\ref{sec:corrections} are modified by their errors to derive the corresponding systematic uncertainty of \CZ. In addition, the impact of the measured SF dependence of the final \CZ~value on the event centrality is also evaluated. The total relative uncertainty from these two sources ranges from 3.1\% at midrapidity ($|y|<0.5$) to 4.5\% at forward \Zboson boson rapidities and gives a contribution, constant as a function of the event centrality, of $\sim$3.4\% for \Zboson boson yields. \par 
 
The main contribution to the systematic uncertainty in the electron channel comes from the uncertainties in measuring the reconstruction efficiency scale factor.  Uncertainties related to this efficiency are classified as either correlated or uncorrelated, and are propagated accordingly to the final measurement uncertainty. The correlated uncertainty component of the SF is obtained by varying the requirements on the tag electron identification and isolation and on the invariant mass of the tag-and-probe pair. The statistical, uncorrelated, components of the scale factor uncertainties are propagated to the measurements via MC pseudo-experiments, while the systematic components are propagated as a single variation fully correlated across all electron $\eta$ intervals. This source gives a 2.5--5\% uncertainty as a function of rapidity and around 3\% for all centrality intervals. \par
 
The effect of the calibration and energy scale correction uncertainty of electrons and muons is negligible.

An additional uncertainty of the bin-by-bin correction is due to the parameterisation of the rapidity and centrality dependence of \CZ\ described in Eq.~(\ref{eq:Z_eff}) and it stems primarily from the statistical uncertainty of the MC data set. To estimate uncertainties associated with these assumptions the parameters of the function $G(y, \sumfcalet)$ are varied by the errors of the parameters of the parabolic fit including covariance between the parameters. The data are corrected with these variations, and the difference between these results and the standard correction are taken as an estimate of the systematic uncertainty. In the muon channel the uncertainty associated with this source ranges from 0.4\% to 1.4\% in rapidity bins and is constant at $\sim$0.5\% as a function of centrality, with the exception of the most peripheral bin where the uncertainty is $\sim$1\%. In the electron channel, the uncertainty as a function of the rapidity ranges from 0.4\% to 1.6\% and is $\sim$1\% for most centrality intervals, although in the most peripheral bin this contribution rises to $\sim$2\%. \par

For both channels and the combined result, the uncertainties of the geometric parameters (\avgTaa and \avgNpart) listed in Table~\ref{table:centrality} range from about 1\% in central collisions to about 12\% in peripheral collisions. These are treated as fully correlated between the channels. 
Finally, the total uncertainty for the \pp measurement~\cite{pp_paper_zw2019} used for the \RAA calculation is 2.3\%. 
 
\section{Results \label{sec:results}}
\begin{figure}[tbp]
\centering
\includegraphics[width=0.48\textwidth]{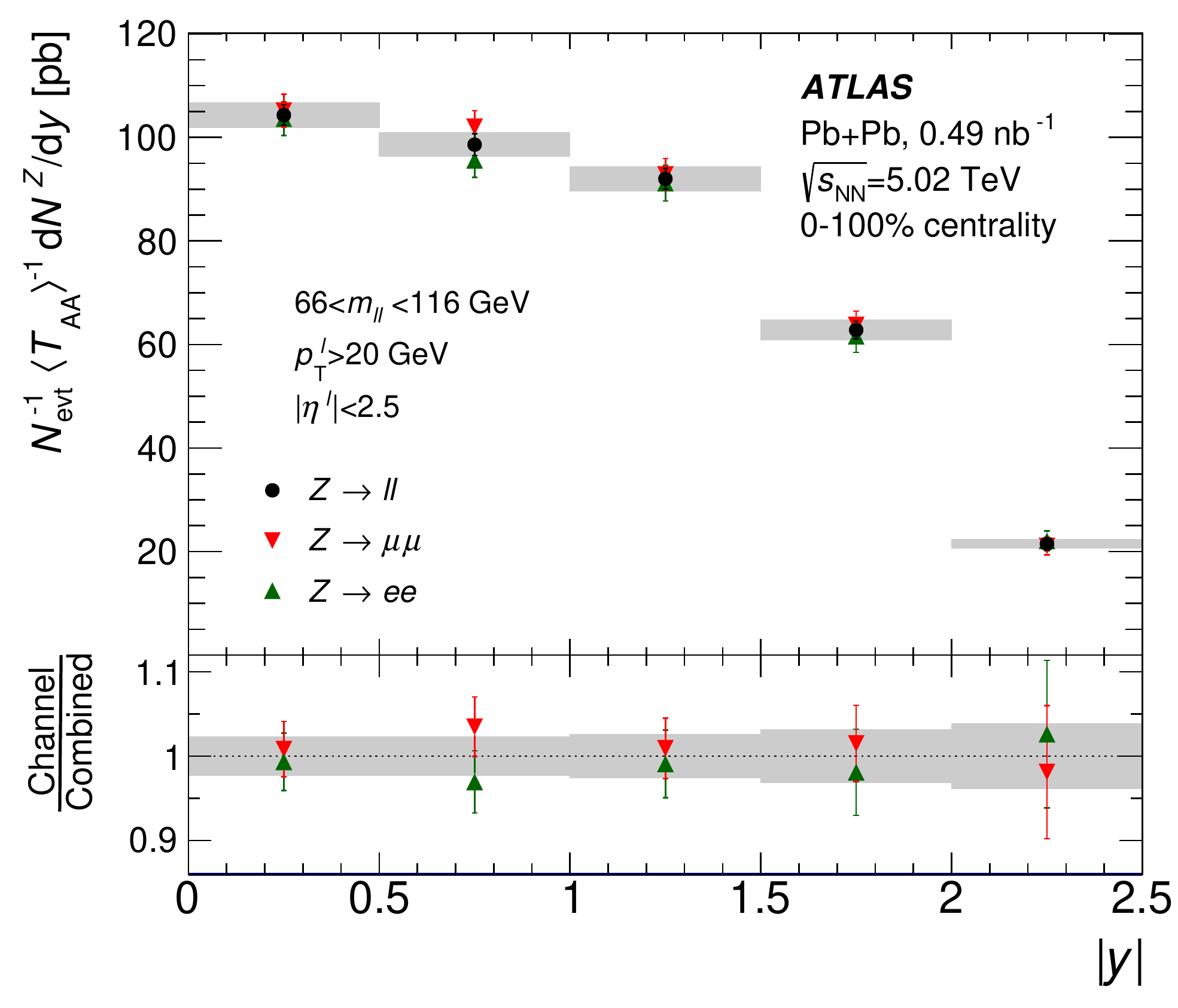}
\includegraphics[width=0.48\textwidth]{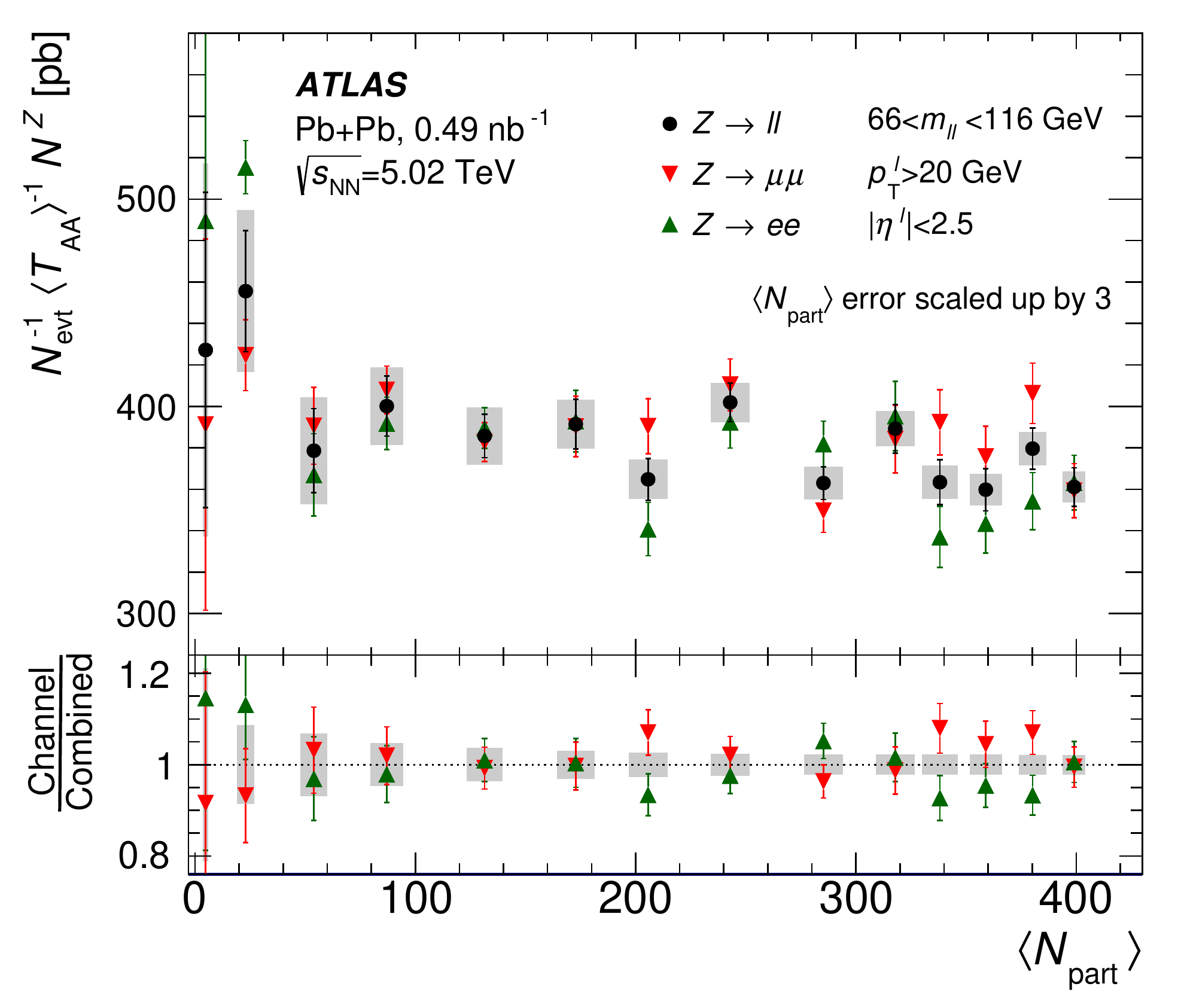}
\caption{Normalised \Zboson boson yields measured in the muon and electron decay channels together with the combined yield as a function of (left) rapidity and (right) \avgNpart. Lower panels show the ratio of individual channels to the combined result. The error bars in the upper panels show the total uncertainty for muons and electrons and the statistical uncertainty for the combined data. In the lower panels, the error bars show the statistical uncertainty. The shaded band (left) and boxes (right) show the systematic uncertainty of the combined result in both panels.
The width of each error box in the right panel corresponds to the systematic uncertainty of \avgNpart, scaled by a factor of three for clarity.
}
\label{fig:comb1}
\end{figure}

The rapidity distributions of the $Z$ boson yield for the muon and electron decay channels, normalised by the number of MB events and mean nuclear thickness function \avgTAA, are shown in the upper-left panel of Figure~\ref{fig:comb1}.
The right panel of the figure shows the \avgNpart dependence of the normalised \Zboson boson yield in the fiducial acceptance where the systematic uncertainties of the \avgNpart values are scaled by a factor of three for clarity. The measurements performed in the two channels are combined using the Best Linear Unbiased Estimate (BLUE) method~\cite{Valassi:2003mu}, accounting for the correlations of the systematic uncertainties across the channels and measurement bins. The combined result is shown in Figure~\ref{fig:comb1} together with the combined statistical and systematic uncertainties. The level of agreement between the channels shown in the lower panels of the figure is quantified as $\chi^{2}/N_{\textrm{dof}}=1.7/5$ as a function of rapidity and $\chi^{2}/N_{\textrm{dof}}=21.6/14$ as a function of centrality.
 
The measured \Zboson boson yields are compared with theoretical predictions obtained using a modified version of {\textsc DYNNLO}~1.5~\cite{Catani:2007vq,Catani:2009sm} optimised for fast computations. The
calculation is performed at $O(\alphas)$ in QCD and at leading order in the EW theory, with parameters set according to the $G_\mu$ scheme~\cite{Hollik:1988ii}. The input parameters (the
Fermi constant 
$G_\text{F}$,
the masses and widths of \Wboson\ and \Zboson\ bosons, and the CKM matrix elements) are taken from Ref.~\cite{Agashe:2014kda}.
The {\textsc DYNNLO} predictions are calculated using the free proton PDF set CT14 NLO~\cite{Dulat:2015mca} typically used to compare with the $pp$ data and, additionally, the nuclear PDF sets nCTEQ15 NLO~\cite{Kovarik:2015cma} and EPPS16 NLO~\cite{Eskola:2017rmp}, which are averaged over each Pb nucleus.
In addition, the parton-level NLO prediction from the MCFM code~\cite{Campbell:2015qma}, interfaced to the CT14 NLO PDF set, is calculated. This takes into account the isospin effect, due to different partonic compositions of protons and neutrons in the Pb nuclei, which is neglected in the DYNNLO calculations.
The renormalisation and factorisation scales, respectively denoted by $\muR$ and $\muF$, are set to the value of lepton pair invariant mass. The uncertainties of these predictions are derived as follows. The effects of PDF uncertainties are evaluated from the variations corresponding to each NLO PDF set. Uncertainties due to the scales are defined by the envelope of the variations obtained by changing $\muR$ and $\muF$ by a factor of two from their nominal values and imposing $0.5\leq \muR/\muF \leq2$. The uncertainty induced by the strong coupling constant is estimated by varying $\alphas$ by $\pm 0.001$ around the central value of $\alphas(m_Z) = 0.118$,
following the prescription of Ref.~\cite{Dulat:2015mca}; the effect of these variations is estimated by comparing the {\textsc CT14nlo\_as\_0117} and {\textsc CT14nlo\_as\_0119} PDF sets~\cite{Dulat:2015mca} to {\textsc CT14 NLO}.  Imperfect knowledge of the proton PDF and the scale variations are the main contributions to the total theory uncertainties. In calculating the \RAA predictions, only the nuclear PDF uncertainties contribute since the CT14 NLO uncertainties cancel.
 
\begin{figure}[tbp]
\centering
\includegraphics[width=0.48\textwidth]{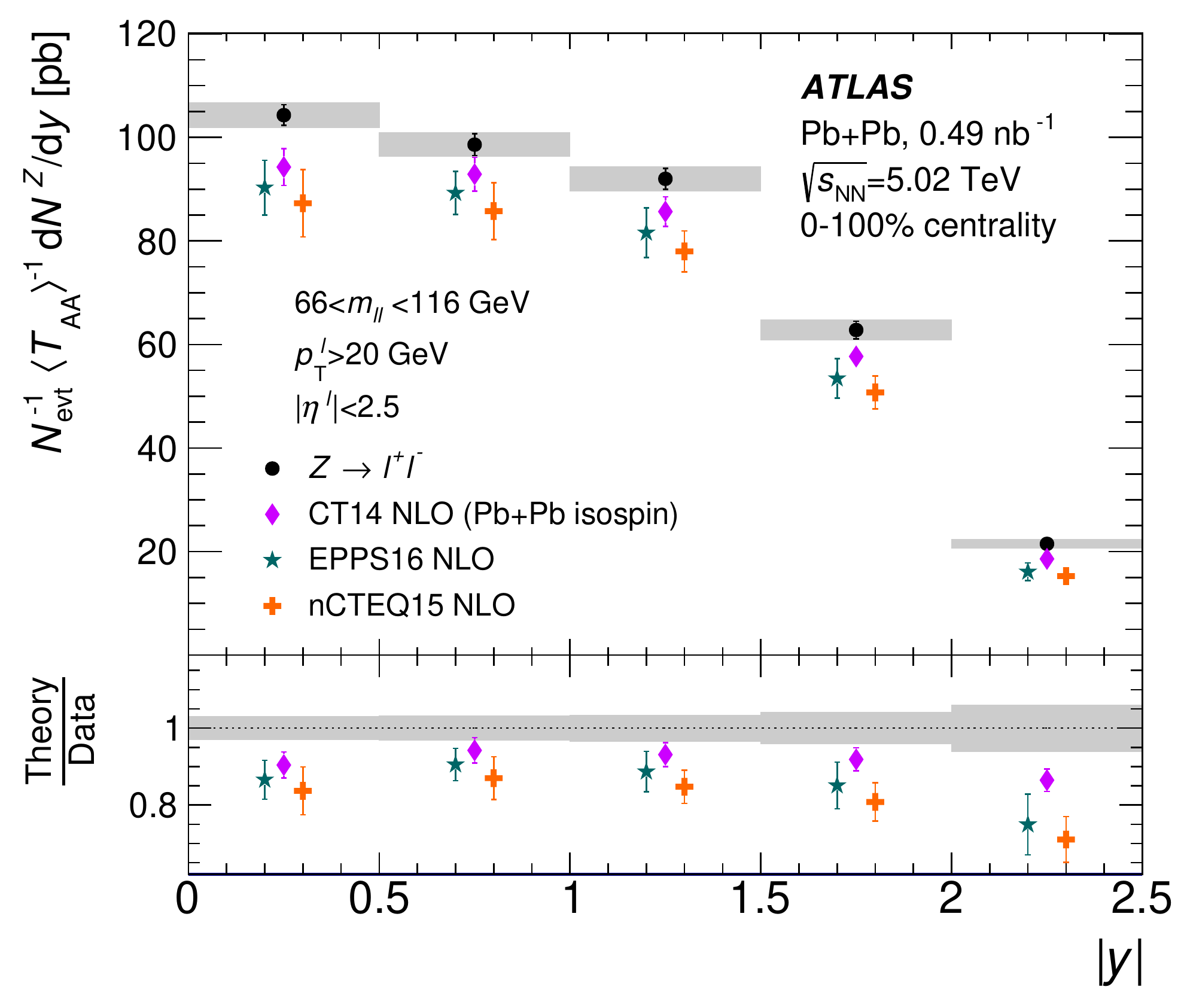}
\includegraphics[width=0.48\textwidth]{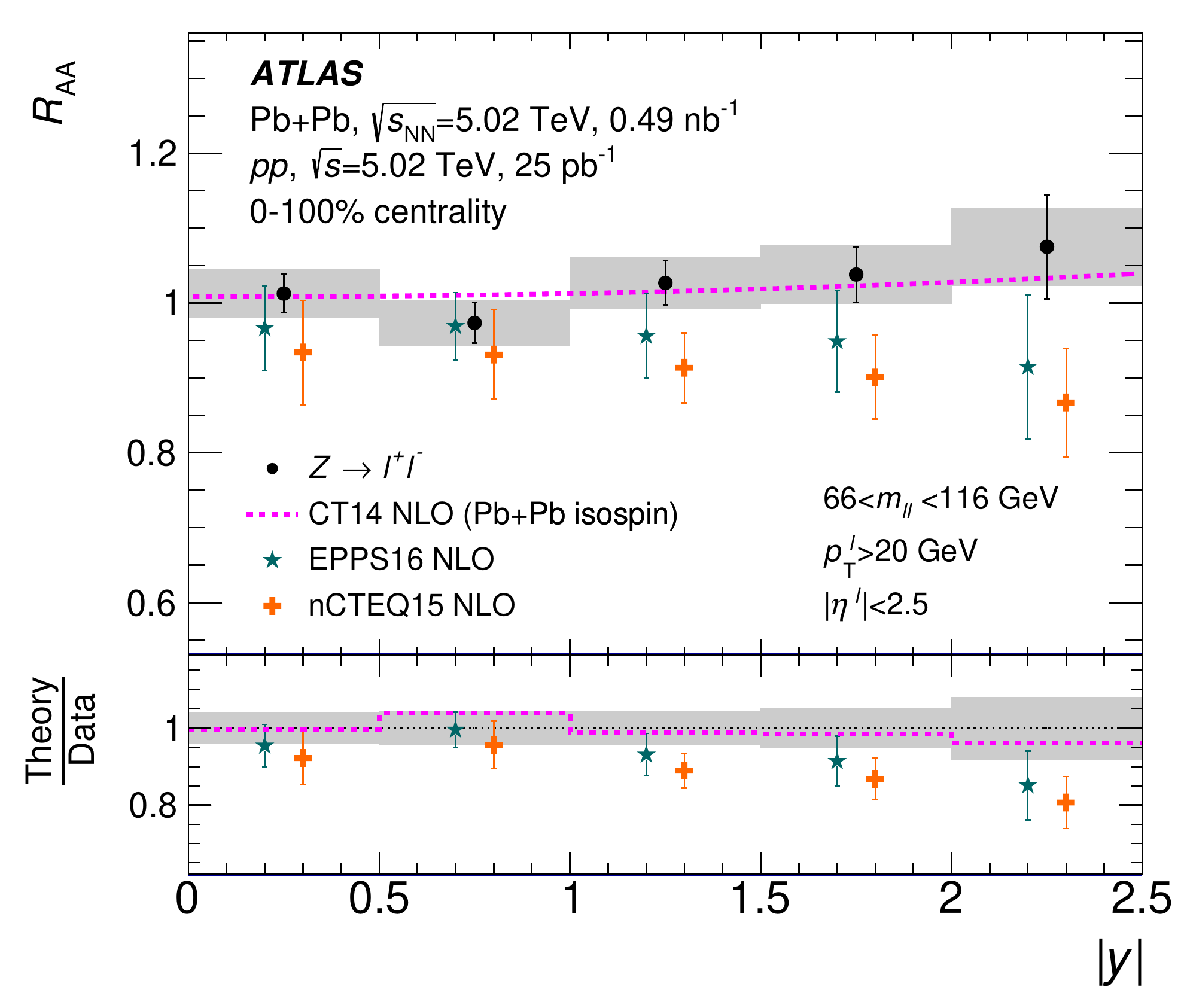}
\caption{The upper panels show the rapidity dependence of (left) the normalised \Zboson boson yields and (right) of the \RAA compared with theoretical predictions. The lower panels show the ratio of the theoretical predictions to the data. The expected contribution of the isospin effect to the \RAA is shown in the upper-right panel by the dashed line.
The error bars on the data points indicate the statistical uncertainties and the shaded boxes show the systematic uncertainties.
The error bars on predictions show the theoretical uncertainty. The points corresponding to nuclear PDF predictions are shifted horizontally relative to the bin centre for clarity.}
\label{fig:raa_absy}
\end{figure}

In Figure~\ref{fig:raa_absy} the normalised \Zboson boson yield is compared between the combined measurement and the theoretical predictions calculated with the CT14, nCTEQ15 and EPPS16 NLO PDF sets, with uncertainties assigned as previously described.
\begin{figure}[tbp]
\centering
\includegraphics[width=0.48\textwidth]{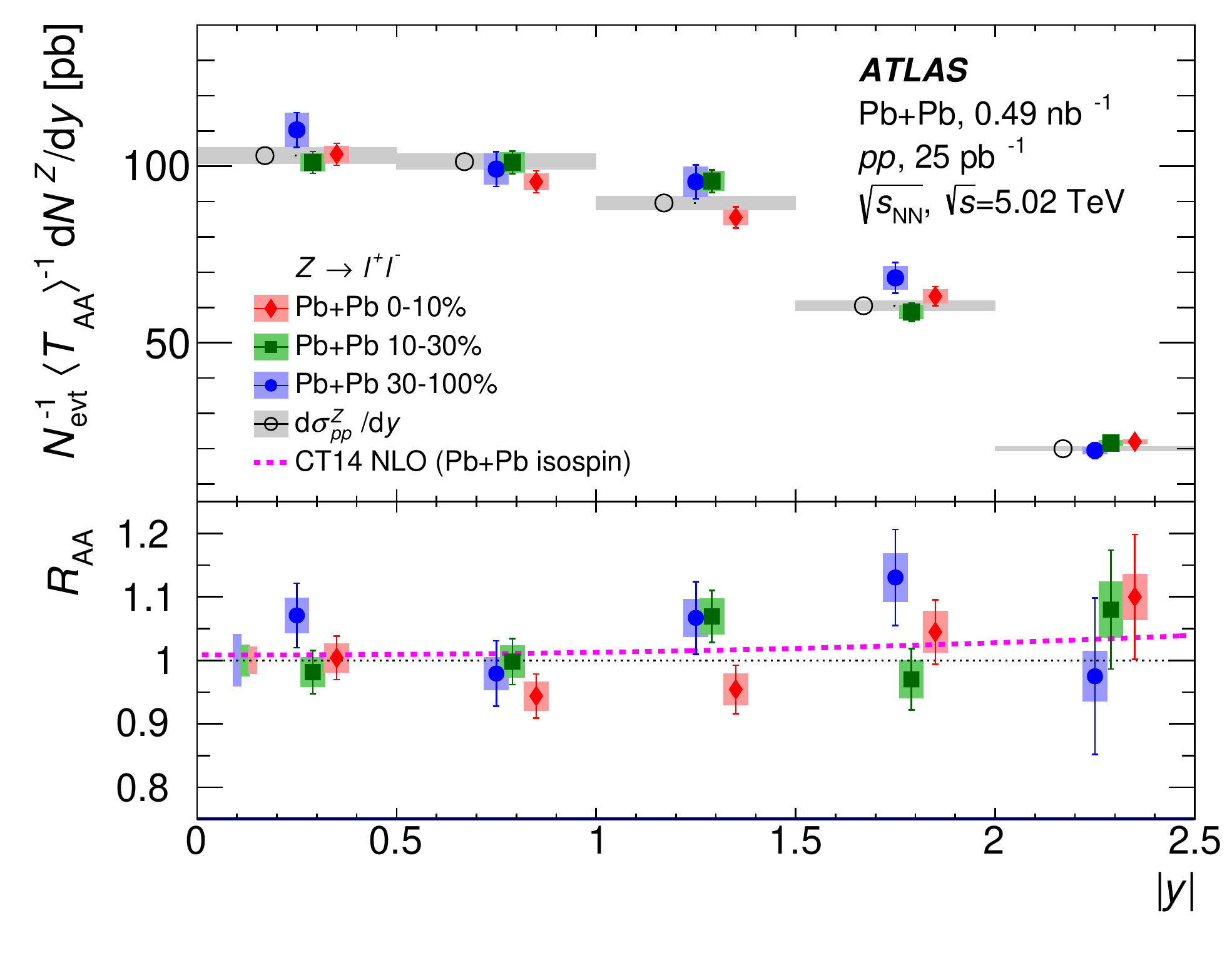}
\caption{Normalised \Zboson boson yield versus rapidity measured in three centrality intervals: peripheral 30--100\% (blue circles), mid-central 10--30\% (green squares) and central 0--10\% (red diamonds).
The differential $Z$ boson cross sections measured in $pp$ collisions are shown by open circles~\cite{pp_paper_zw2019}. The error bars on the data points indicate the statistical uncertainties and the shaded boxes show the total systematic uncertainties.
The lower panel shows the \RAA and the contribution from the isospin effect calculated with CT14 NLO PDF (dashed line). The shaded boxes at unity indicate the combined uncertainty from the \pp data added in quadrature to the \TAA uncertainty. The points in each centrality interval are shifted horizontally relative to the bin centre for clarity.}
\label{fig:raa_absy_cent}
\end{figure}
All calculations lie 1--3$\sigma$ below the data in all rapidity intervals, integrated over event centrality. Calculations using nuclear PDF sets deviate from the data more strongly than calculations based only on the CT14 NLO PDF set. A similar observation was made with the \pp collision system~\cite{pp_paper_zw2019} where systematic deviations from the measured values are observed for calculations made at the NNLO. When comparing the measured \RAA with calculations, shown in the right panel of Figure~\ref{fig:raa_absy}, residual deviations from the data are observed. The trend observed in data is consistent with the isospin effect only, expected from the different valence quark content of protons and neutrons in the Pb nucleus, shown in the upper-right panel of Figure~\ref{fig:raa_absy} by the dashed line.

Figure~\ref{fig:raa_absy_cent} shows the normalised \Zboson boson yield as a function of rapidity for three centrality intervals. The results are consistent with each other within their respective statistical uncertainties. The size of the current data sample precludes making a more definitive statement about any possible modification of the \Zboson boson rapidity distribution with centrality.

\begin{figure}[tbp]
\centering
\includegraphics[width=0.53\textwidth]{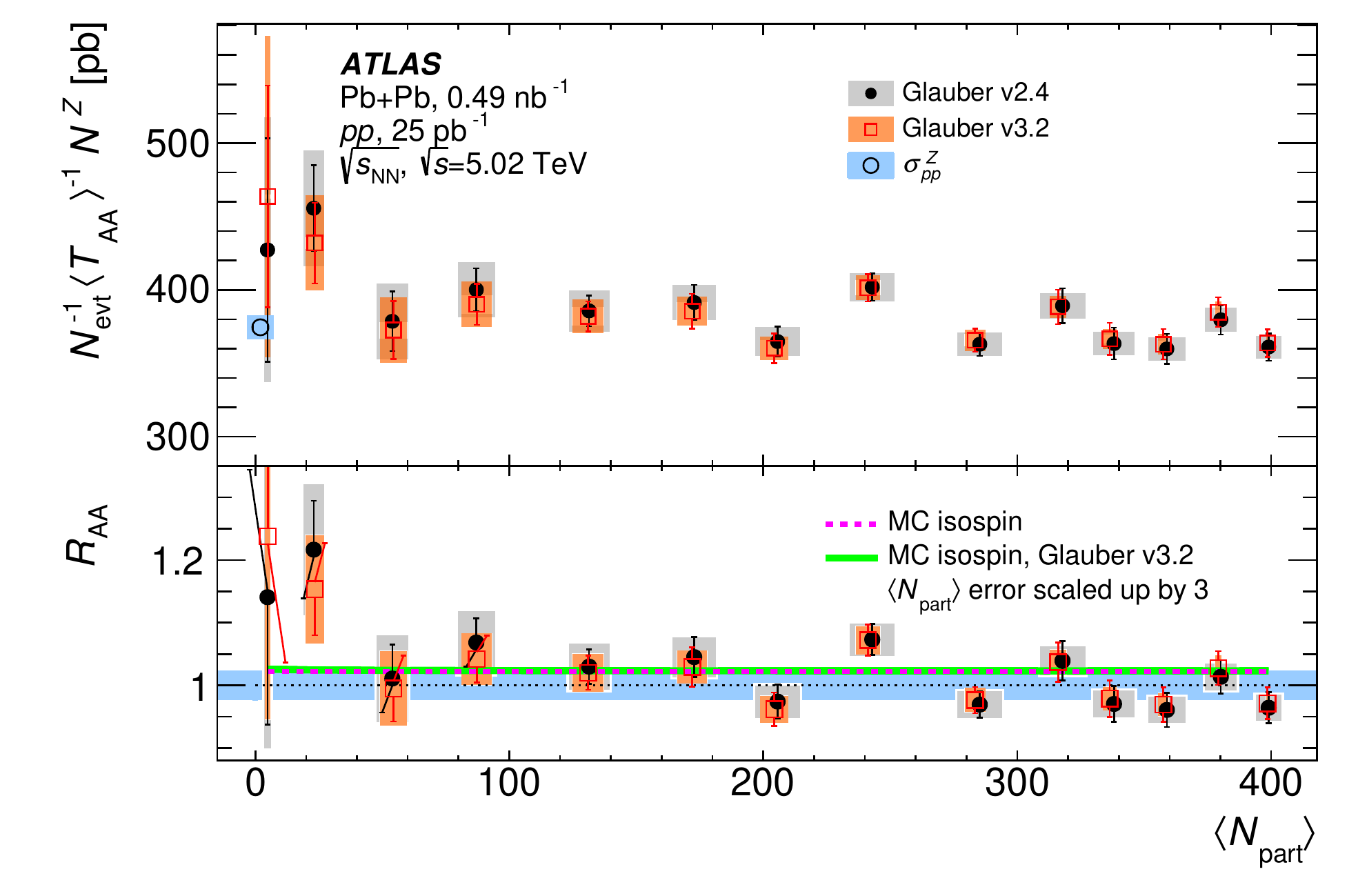}
\caption{The upper panel shows the normalised \Zboson boson yield as a function of \avgNpart. The integrated fiducial cross section measured in the $pp$ system is shown at $\Npart=2$~(open circle). The lower panel shows the nuclear modification factor. The error bars on the data points indicate the statistical uncertainties and the shaded boxes show the total systematic uncertainties. The blue shaded band indicates the total uncertainty of the \pp data. The figure shows the results calculated with Glauber MC v2.4~\cite{Miller:2007ri} and v3.2~\cite{Loizides:2017ack}. 
The dashed and the full line indicate CT14 NLO PDF calculations that account for isospin effects for the two Glauber MC versions.
The width of the error box for each centrality interval corresponds to the systematic uncertainty in \avgNpart, scaled up by a factor of three for clarity.
}
\label{fig:raa_npart}
\end{figure}

Figure~\ref{fig:raa_npart} shows the centrality dependence of the normalised \Zboson boson yield and of \RAA. 
The point corresponding to the \pp cross section~\cite{pp_paper_zw2019} is shown in the plot at $\Npart=2$.
The results are derived from Glauber MC v2.4 and a newer version v3.2 following the same procedure as described in Ref.~\cite{Aad:2019sfe}. The results are found to be consistent with each other within experimental uncertainties.
The new Glauber MC calculation~\cite{Loizides:2017ack} implements a more advanced treatment of the nuclear density profile and an updated experimental value of the nucleon--nucleon inelastic cross section. From these two updates, the former directly affects the normalised $Z$ boson yield derived in this analysis while the updated cross section value has no appreciable effect. The new model also leads to a set of reduced systematic uncertainties compared with the previous version.
 
In the estimation of the isospin effect contribution shown with dashed and full lines in Figure~\ref{fig:raa_npart}, the Glauber MC v3.2 model accounts for the slightly larger radius of the neutron distribution compared to the protons in the Pb nucleus, often called the neutron skin effect. 
However, due to the weak dependence of $Z$ boson production on the isospin content of the colliding baryons, the predictions of the two Glauber MC versions give essentially the same result.
 
The normalised yields are consistent with the \pp cross section at all measured centralities and show only a weak dependence on \Npart. The values of \RAA, shown in the lower panel, are consistent with unity within the total uncertainty.
When the isospin effect is taken into account as shown with the dashed line, the model seems to agree better with the data at low \Npart values rather than at high values. 
To quantify the dependence of \RAA on \Npart, the data are fit to a linear function. Including statistical and systematic uncertainties the decrease in the \RAA value between the most-peripheral (80--100\%) and most-central (0--2\%) centrality intervals is found to be $(10 \pm 7)\%$ and $(6 \pm 6)\%$ for Glauber MC v2.4 and v3.2, respectively.

\begin{figure}[tbp]
\centering
\includegraphics[width=0.48\textwidth]{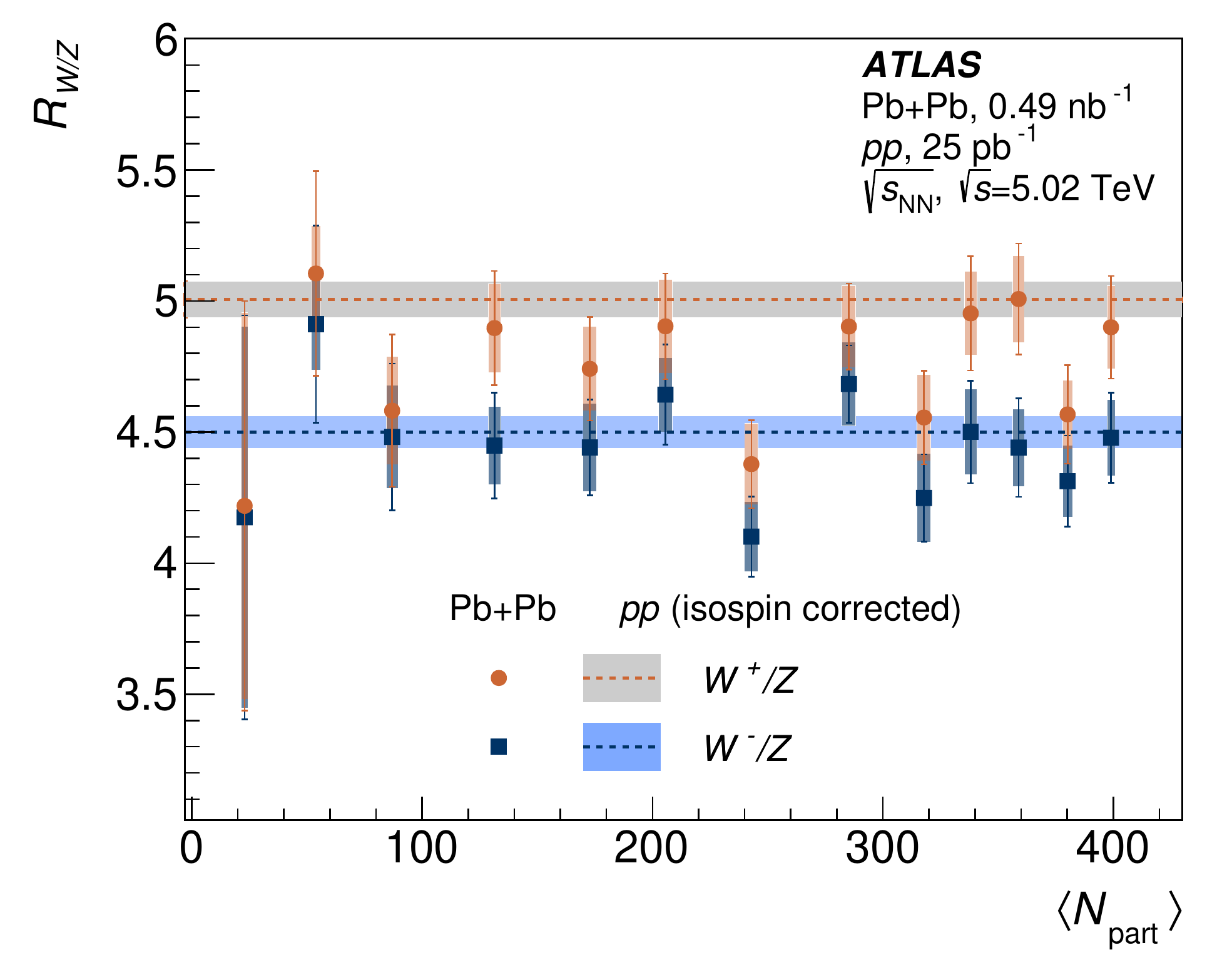}
\caption{Ratios of $W^{+}$ (circles) and $W^{-}$ (squares) yields measured in the same data set~\cite{Aad:2019sfe} to the yields of \Zboson bosons versus $\langle \Npart \rangle$ compared with the ratios measured in \pp data~\cite{pp_paper_zw2019} (shaded bands) scaled by isospin factors obtained from the CT14 NLO PDF calculation. The error bars on the data points indicate the statistical uncertainties and the shaded boxes show the total systematic uncertainties. }
\label{fig:ratio_wz}
\end{figure}

The \Zboson boson measurement is used to compare the \Npart dependence of \Wboson~\cite{Aad:2019sfe} and \Zboson boson production by calculating their yield ratios as shown in Figure~\ref{fig:ratio_wz}, where the uncertainties of the two measurements are conservatively treated as uncorrelated.
The data points are compared with the ratio measured in \pp collisions~\cite{pp_paper_zw2019} that is scaled by the isospin factors calculated using the CT14 NLO PDF set. The measurements for both channels are found to be consistent with the scaled \pp measurement and show a constant behaviour as a function of centrality.
 
The trend of the points shown in Figure~\ref{fig:raa_npart} for \Zboson bosons is different from the trend observed by the ALICE Collaboration in the measurements with charged hadrons with high transverse momentum~\cite{Acharya:2018njl}.
It was recently shown in Ref.~\cite{Morsch:2017brb}, that the $R_\mathrm{AA}$ in peripheral nucleus--nucleus collisions can deviate from unity due to a biased classification of the event geometry for events containing a hard process.
In that analysis, the value of $R_\mathrm{AA}$ without any nuclear effects was determined by using the HG-\textsc{Pythia} model~\cite{Morsch:2017brb}, which can create an ensemble of events where the \textsc{Hijing}~\cite{Wang:1991hta} event generator is used to determine the number of hard sub-interactions for each event, and the particle production is determined solely by superimposing a corresponding number of \textsc{Pythia} 6.4~\cite{Sjostrand:2006za} events.
The model is able to qualitatively explain the ALICE measurement in the peripheral region.
 
\begin{figure}[tbp]
\centering
\includegraphics[width=0.48\textwidth]{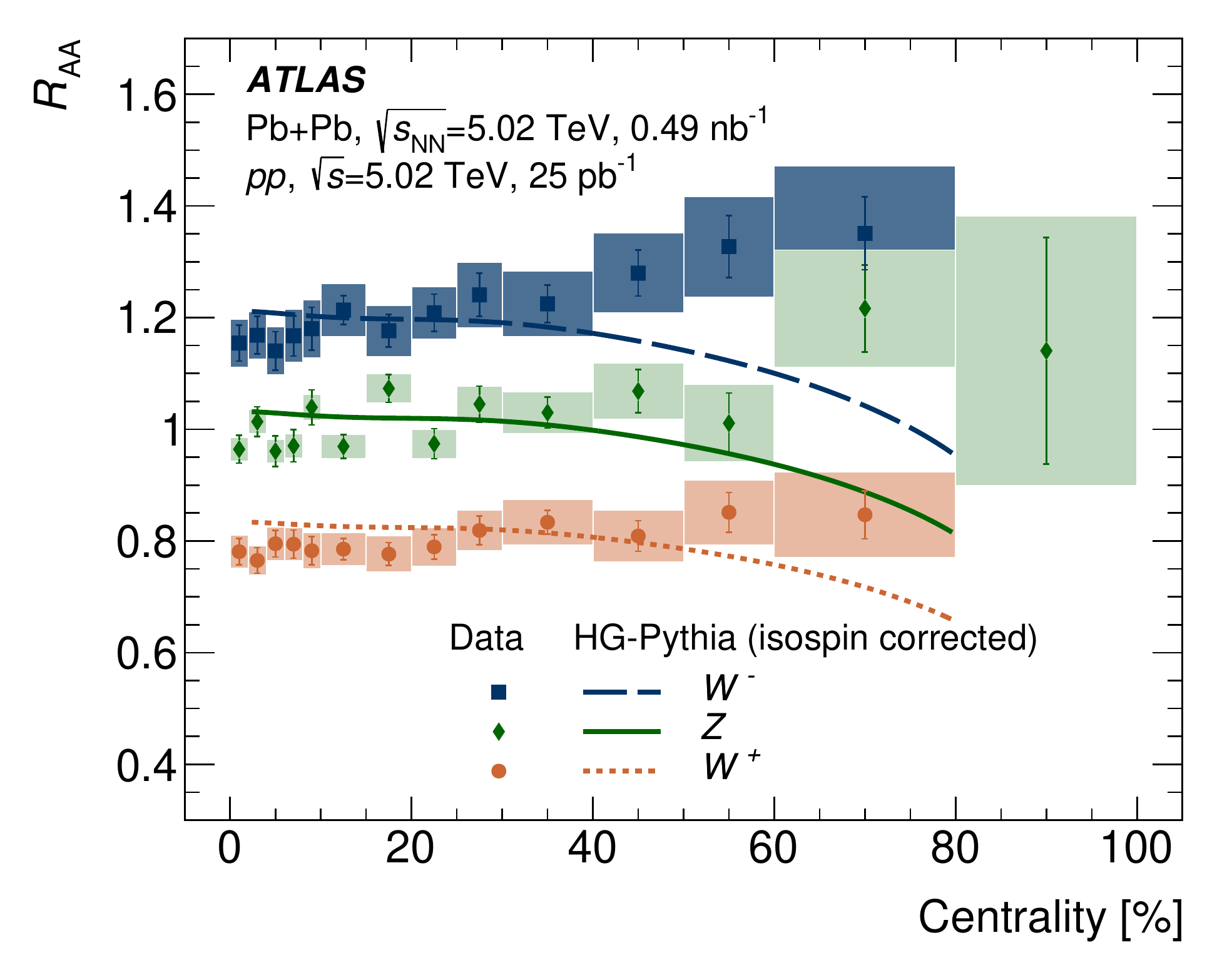}
\caption{Nuclear modification factor \RAA in centrality intervals compared with the HG-\textsc{Pythia} model~\cite{Morsch:2017brb} scaled by the isospin factors obtained from the CT14 NLO PDF calculation. The error bars on the data points indicate the statistical uncertainties and the shaded boxes show the total systematic uncertainties. The model accounts for a biased classification of the event geometry for events containing a hard process. The bins are ordered in centrality percentile, starting from central events on the left towards more peripheral on the right. The $Z$ boson measurement extends to the most peripheral 80--100\% centrality interval.
The comparison is shown in the 0--80\% centrality interval where the different smearing of the ATLAS centrality estimator (\sumfcalet) is found to have a negligible effect.
}
\label{fig:bias_raa}
\end{figure}

Figure~\ref{fig:bias_raa} shows the \RAA for $\Wboson^{\pm}$~\cite{Aad:2019sfe} and \Zboson bosons compared with the HG-\textsc{Pythia} model. 
To compare the model with the measurements of massive electroweak bosons the results are corrected for the isospin effect using the CT14 NLO PDF. All three data sets show trends that are consistent between the species, but that differ from their corresponding model predictions. This suggests that the apparent suppression mechanism~\cite{Morsch:2017brb} that explains the ALICE data~\cite{Acharya:2018njl} for the yields of high-\pT charged particles does not have the same effect on the yields of massive electroweak bosons.

\section{Summary}\label{sec:summary}
The \Zboson boson production yield per minimum-bias collision, scaled by the mean nuclear thickness function \avgTaa is reported in \PbPb collisions at a nucleon--nucleon centre-of-mass energy $\sqn=5.02$~\TeV. The measurement is based on data taken by the ATLAS detector at the LHC corresponding to an integrated luminosity of 0.49\;\inb.  Normalised yields are reported in the electron and muon decay channels, differentially in rapidity and collision centrality in the mass window $66<\mll<116$~\GeV.  The fiducial region is defined using the lepton kinematics and detector acceptance.  The electron channel and muon channel results are found to agree within the measurement precision and are combined for the final result. \par
 
The normalised $Z$ boson yields measured in \PbPb collisions are 1--3$\sigma$ higher than NLO pQCD predictions with both free and nuclear PDF sets, where the difference increases towards forward $Z$ boson rapidity. The nuclear modification factor is measured differentially as a function of $Z$ boson rapidity and event centrality. It is found to be consistent with unity in centrality and to agree with the prediction based on the CT14 PDF set that takes isospin into account.
This behaviour is also consistent with the ATLAS measurement performed with $\Wboson$ bosons. The yield ratios $\Wboson/\Zboson$ are found to be constant as a function of centrality within the uncertainties of the measurements. Unlike high-\pT charged hadrons measured by the ALICE Collaboration, $W$ and $Z$ bosons show no indication of yield suppression in peripheral collisions. 
 
 
\section*{Acknowledgements}
 
We thank CERN for the very successful operation of the LHC, as well as the
support staff from our institutions without whom ATLAS could not be
operated efficiently.
 
We acknowledge the support of ANPCyT, Argentina; YerPhI, Armenia; ARC, Australia; BMWFW and FWF, Austria; ANAS, Azerbaijan; SSTC, Belarus; CNPq and FAPESP, Brazil; NSERC, NRC and CFI, Canada; CERN; CONICYT, Chile; CAS, MOST and NSFC, China; COLCIENCIAS, Colombia; MSMT CR, MPO CR and VSC CR, Czech Republic; DNRF and DNSRC, Denmark; IN2P3-CNRS, CEA-DRF/IRFU, France; SRNSFG, Georgia; BMBF, HGF, and MPG, Germany; GSRT, Greece; RGC, Hong Kong SAR, China; ISF and Benoziyo Center, Israel; INFN, Italy; MEXT and JSPS, Japan; CNRST, Morocco; NWO, Netherlands; RCN, Norway; MNiSW and NCN, Poland; FCT, Portugal; MNE/IFA, Romania; MES of Russia and NRC KI, Russian Federation; JINR; MESTD, Serbia; MSSR, Slovakia; ARRS and MIZ\v{S}, Slovenia; DST/NRF, South Africa; MINECO, Spain; SRC and Wallenberg Foundation, Sweden; SERI, SNSF and Cantons of Bern and Geneva, Switzerland; MOST, Taiwan; TAEK, Turkey; STFC, United Kingdom; DOE and NSF, United States of America. In addition, individual groups and members have received support from BCKDF, CANARIE, CRC and Compute Canada, Canada; COST, ERC, ERDF, Horizon 2020, and Marie Sk{\l}odowska-Curie Actions, European Union; Investissements d' Avenir Labex and Idex, ANR, France; DFG and AvH Foundation, Germany; Herakleitos, Thales and Aristeia programmes co-financed by EU-ESF and the Greek NSRF, Greece; BSF-NSF and GIF, Israel; CERCA Programme Generalitat de Catalunya, Spain; The Royal Society and Leverhulme Trust, United Kingdom.
 
The crucial computing support from all WLCG partners is acknowledged gratefully, in particular from CERN, the ATLAS Tier-1 facilities at TRIUMF (Canada), NDGF (Denmark, Norway, Sweden), CC-IN2P3 (France), KIT/GridKA (Germany), INFN-CNAF (Italy), NL-T1 (Netherlands), PIC (Spain), ASGC (Taiwan), RAL (UK) and BNL (USA), the Tier-2 facilities worldwide and large non-WLCG resource providers. Major contributors of computing resources are listed in Ref.~\cite{ATL-GEN-PUB-2016-002}.
 

\printbibliography

\clearpage 
 
\begin{flushleft}
{\Large The ATLAS Collaboration}

\bigskip

G.~Aad$^\textrm{\scriptsize 101}$,    
B.~Abbott$^\textrm{\scriptsize 128}$,    
D.C.~Abbott$^\textrm{\scriptsize 102}$,    
A.~Abed~Abud$^\textrm{\scriptsize 70a,70b}$,    
K.~Abeling$^\textrm{\scriptsize 53}$,    
D.K.~Abhayasinghe$^\textrm{\scriptsize 93}$,    
S.H.~Abidi$^\textrm{\scriptsize 167}$,    
O.S.~AbouZeid$^\textrm{\scriptsize 40}$,    
N.L.~Abraham$^\textrm{\scriptsize 156}$,    
H.~Abramowicz$^\textrm{\scriptsize 161}$
% ...
\bigskip
\\

% ...
$^{40}$Niels Bohr Institute, University of Copenhagen, Copenhagen; Denmark.\\
% ...
$^{53}$II. Physikalisches Institut, Georg-August-Universit\"{a}t G\"ottingen, G\"ottingen; Germany.\\
% ...
$^{70}$$^{(a)}$INFN Sezione di Pavia;$^{(b)}$Dipartimento di Fisica, Universit\`a di Pavia, Pavia; Italy.\\
% ...
$^{93}$Department of Physics, Royal Holloway University of London, Egham; United Kingdom.\\
% ...
$^{101}$CPPM, Aix-Marseille Universit\'e, CNRS/IN2P3, Marseille; France.\\
$^{102}$Department of Physics, University of Massachusetts, Amherst MA; United States of America.\\
% ...
$^{128}$Homer L. Dodge Department of Physics and Astronomy, University of Oklahoma, Norman OK; United States of America.\\
% ...
$^{156}$Department of Physics and Astronomy, University of Sussex, Brighton; United Kingdom.\\
% ...
$^{161}$Raymond and Beverly Sackler School of Physics and Astronomy, Tel Aviv University, Tel Aviv; Israel.\\
% ...
$^{167}$Department of Physics, University of Toronto, Toronto ON; Canada.\\
% ...
$^{a}$ Also at Borough of Manhattan Community College, City University of New York, New York NY; United States of America.\\
$^{b}$ Also at CERN, Geneva; Switzerland.\\
% ...

\end{flushleft}


\end{document}